\documentclass[reprint, amsmath,amssymb, aps]{revtex4-1}

\usepackage{graphicx}
\usepackage{dcolumn}
\usepackage{bm}
\usepackage{color}

\begin{document}

\preprint{APS/123-QED}

\title{Constraints on leptophilic light dark matter from internal heat flux of Earth}

\author{Bhavesh Chauhan}
\altaffiliation[Also at ]{Indian Intitute of Technology, Gandhinagar (India)}
\email{bhavesh@prl.res.in}
\author{Subhendra Mohanty}
\email{mohanty@prl.res.in}
\affiliation{ Theoretical Physics Department, Physical Research Laboratory, Ahmedabad (India).}

\date{\today}

\begin{abstract}
 Dark Matter in Earth intersecting orbits can scatter off the electrons and lose energy, and finally be gravitationally bound to Earth. Eventually they lose enough energy and accumulate at the core. It is assumed that DM annihilates/decays predominantly into Standard Model particles inside Earth. The heat flux from these processes is compared with the experimentally measured value of internal heat flux of Earth which is 44 TW. Assuming steady state between capture and annihilation/decay, we put constraints on the scattering cross section of DM with electrons as a function of their mass. For low mass regions ($<10^{-2}$GeV), these constraints on leptophilic DM are better than ones obtained from direct-detection experiments. 
\end{abstract}

\maketitle

\section{Introduction}
	There is evidence for Dark Matter (DM) from rotation curves of galaxies, and Bullet cluster \cite{Garrett}\cite{Profumo}, however there is no firm evidence from direct detection in terrestrial experiments. These experiments search for nuclear recoil from coherent scattering of 10-100 GeV mass DM particles with heavy nuclei to put constraints on scattering cross section. In these experiments, the ionization of electrons is assumed to be from the radioactivity in the background, and only non-ionizing nuclear scatterings are considered as evidence of DM  signal. Consequently, the bounds on leptophilic DM which scatter with the electrons in the atoms are weak \cite{Essig2}. There are well motivated models of leptophilic DM with mass in the sub GeV range \cite{Chu,Essig1,Feng,Lin, Graham} and it will be of interest to put constraints on the scattering cross section from observations. \\

	Fitting of the rotation curve of the milky way with the DM profile leads us to believe that DM has a specific (NFW, Einasto, Isothermal etc.) profile in position space and a Maxwellian distribution in the momentum space. The density of the DM in solar neighborhood, which is relevant for determining the scattering rates in terrestrial experiments, is fixed to be 0.4 GeV/cm$^3$ in all the profiles \cite{Catena}. \\
	
	In this paper we investigate the gravitational capture of leptophilic DM, and its subsequent  annihilation/decay in Earth. DM in earth intersecting orbits can scatter off electrons of the earth and lose energy. If these scattered DM particles are gravitationally captured in an Earth bound orbit, then eventually they can accumulate inside Earth. These DM particles can annihilate/decay into SM particles and impart their energy to the kinetic energy of atoms. Assuming a steady state between accumulation and annihilations/decays, we predict the heat produced in the Earth from this process. The heat flux of earth has been experimentally measured to be around 44 TW \cite{Pollack}. Measurement of geo-neutrinos tell us that 20-25 TW is contributed by the nuclear reactions in the Earth \cite{Gando}. The source for remaining heat output is uncertain. We take the conservative estimate of 20 TW as the upper limit of heating by DM annihilations, and put constraints on scattering cross section and mass of leptophilic DM.\\
	
	The problem of gravitational capture of Dark Matter has been extensively studied in the past \cite{RefAdler, RefKawasaki, RefMitra, RefKrauss, RefMack} and the internal heat flux of Earth has previously been used to constrain the WIMP parameter space \cite{Mack, RefMack}. However, the mass of DM in these studies typically ranges for 1-$10^3$ GeV. When applied to light DM, these techniques do constrain the WIMP-nucleon cross section, but yield uninteresting results. However, interesting parameter space can be probed for light DM if it interacts with the electrons in the atoms rather than the nucleons. In this paper it is shown that novel constraints can be obtained for light leptophilic DM using Earth's internal heat.\\
	
	In Sec. II we calculate the capture rate and the total accretion rate of DM by Earth. In Sec. III we calculate the heat flux by annihilations/decays of DM inside Earth. In Sec. IV we discuss our results and compare with existing bounds. 		

	\section{Capturable phase space}
	
	The phase space for gravitational capture of DM was first worked out in Ref. \cite{Press}. There are some discrepancies that were pointed out in Ref. \cite{Gould} regarding a factor of 2, and motion relative to galactic halo. We redo the calculation here for completeness and include these factors. It is assumed that the galatic Dark Matter is thermal and follows Maxwellian distribution characterised by rms velocity, $\bar{v}= 300$ km/s. The rate of particles coming from a unit surface element located on a sphere of radius R centered at Earth is then given by, \cite{Press}
	\begin{equation}\label{accrate}
		dF = n_\chi \left(\frac{3}{2 \pi \bar{v}^2}\right)^{3/2} \pi v^3 \exp\left(\frac{-3v^2}{2 \bar{v}^2}\right) d \left(\cos^2 \theta \right) dv
	\end{equation} 
	with integration limits $0 < v < \infty , ~ 0<\theta<\pi/2$ \cite{Press}. One can express this in terms of orbital invariants i.e. specific-energy, and specific-angular-momentum given as
	\begin{equation}
	E = \frac{1}{2}v^2, ~~ J = vR \sin \theta 
	\end{equation}
	respectively. In terms of these new variables, the \emph{total} accretion rate is obtained by multiplying  Eq.\eqref{accrate} with $4\pi R^2$, and the Jacobian of transformation. This accretion rate is given by, 
	\begin{equation}\label{totaccrate}
	d\mathcal{F} = 4 \pi^2 n_\chi \left(\frac{3}{2 \pi \bar{v}^2}\right)^{3/2} \exp\left( \frac{-3E}{\bar{v}^2}\right) dE dJ^2
 	\end{equation}
 	
 	So far, we have neglectied the motion of Earth with respect to the Galactic halo. As worked out in \cite{Gould}, this factor is, 
 	 	 \begin{equation}
 	 	 	\xi_\eta = \frac{\pi^{1/2}}{2 \eta} \text{Erf}(\eta);\hspace{.5cm} \eta = \frac{\tilde{v}}{\bar{v}}
 	 	 \end{equation}
    where $\tilde{v} = 220$ km/s is the average speed of Earth in the halo. It can be seen that $\xi_\eta \approx 0.84$ for Earth. We take this factor into account in the accretion rate.\\
 	
 	For brevity, we will use the dimensionless variables 
 	\begin{equation}
 	 x = \frac{E}{GM/R}, ~~ y = \frac{J^2}{GMR}
 	\end{equation}
  	where M, and R are the mass, and radius of earth respectively. G is the universal gravitational constant. Using the expression of surface escape volocity $w = \sqrt{2GM/R}$, we can express Eq.\eqref{totaccrate} as,  
	\begin{equation}\label{xyaccrate}
		d\mathcal{F} = \xi_\eta \sqrt{\pi} \left(\frac{3}{2}\right)^{3/2}  n_\chi \frac{w^4}{\bar{v}^3} R^2
		\exp\left( \frac{-3 w^2}{2 \bar{v}^2} x\right) dx dy
	\end{equation} 	
 	 
	To determine what fraction of these particles is captured by Earth, we must determine region of the $\left( x,y\right)$ parameter space which kinematically scatters into bound orbits. This is achieved by imposing the following conditions, (i) the perigee of orbit intersects the Earth and, (ii) the particles are trapped by gravitational influence of Earth.

	\subsection{Perigee Constraint}
	The first condition we impose on the particles is for their trajectories should intersect Earth. For Keplerian orbits, the distance of closest approach is given as,
	\begin{equation}
	r_{peri} = \left(\frac{J^2}{GM}\right) \Biggm/ \left(1 + \sqrt{1 + 2 \frac{J^2}{GM}\frac{E}{GM}}\right).
	\end{equation} 
	
	To ensure that the incident DM particle travels a distance of one Earth radius ($R_E$) inside Earth, we require that $r_{peri} \leq 0.86 R_E$. In terms of the dimensionless variables, this is equivalent to, 
	\begin{equation}
	y \leq 1.48 x + 1.72
	\end{equation}

	This appears as an upper limit of integration in the x-y parameter space.
	 
	\subsection{Gravitational Trapping by Earth}
	
	Having restricted our DM particles to the ones that intersect Earth, now we look at the number of times these particles interact with the electrons of Earth in crossing $R_E$. The mean free path between two scatterings is given by, 
	\begin{equation}
	\lambda_\chi = \frac{1}{\sigma_{\chi e} n_e} 
	\end{equation}
	where $n_e$ is the number density of electrons in Earth. To estimate this number, we assume net electric charge neutrality, and thus the number of electrons and protons in Earth is same. The proton to neucleon fraction is estimated by considering the composition of Earth which is 32.1 \% Iron, 30.1\% Oxygen, 15.1\% Silicon, 13.9\% Magnesium, 2.9\% Sulphur, 1.8\% Nickel, 1.5\% Calcium, 1.4\% Aluminium, and other trace elements \cite{Morgan}. This results in, 
	\begin{equation}
	 n_e = 0.48 \frac{M_E}{M_p} \times \frac{1}{\frac{4}{3}\pi R^3} = 1.6 \times 10^{30} \text{m}^{-3}
	\end{equation}
	where $M_E = 6 \times 10^{24}$kg is the mass of Earth. The typical number of scattering can be approximated by, 
	\begin{equation}
	N_{s} \approx \frac{R_E}{\lambda_\chi} = R_E \times \sigma_{\chi e} \times n_e \equiv \frac{\sigma_{\chi e}}{\sigma_{crit}} .
	\end{equation}
	
	For Earth, $\sigma_{crit} \approx 9.5 \times 10^{-38} \text{m}^2$. When $\sigma_{\chi e} > \sigma_{crit}$, DM will definitely scatter from the electrons. But, when $\sigma_{\chi e} < \sigma_{crit}$, the probablity of scattering is proportional to $N_s$ \cite{Press}. This constant of proportiality was evaluated in \cite{Press} to be of order unity and can be ignored for sake of approximation. The probability of scattering can thus be writted in the form,
	\begin{equation}\label{probscat}
	g(\sigma) = \theta(\sigma - \sigma_{crit}) + \frac{\sigma}{\sigma_{crit}}\theta(\sigma_{crit} - \sigma)
	\end{equation}  
	where $\theta(x)$ is the Heaviside Theta function. 
	
	After $N_s$ scatterings, the DM particle must remain in the gravitational influence of earth. This, along with the perigee constraint, guarantees that the particle will re-renter Earth's surface and subsequently lose more energy via scatterings. Eventually it will drift to the core of earth where it annihilates or decays. \\
	
	The energy of DM particle after one scattering is approximated by, \cite{Mack}
	\begin{equation}
	E_f \approx E_i \times \left(1 - \frac{2 m_\chi m_e}{\left(m_\chi + m_e \right)^2}\right)
	\end{equation}
	where the term in parenthesis will be subsequently represented by $f(\mu = m_\chi/m_e)$. After $N_s$ such scatterings, the final specific energy of the DM particle will be
	\begin{equation}\label{EF}
	 E_f = f(\mu)^{N_s} E_i. 
	\end{equation}
	After each scattering, apogee of the DM orbit decreases. If the apogee is outside the Hill spehere of Earth, then there is a possibility that the particle may be captured by the Sun. For an astronomical object, Hill sphere is the region in which it dominates the attraction of satellites. Hence we impose a condition that the apogee of scattered orbit must be inside the Hill sphere of Earth which is calculated as,  
	\begin{equation}
	 R_H = a \left(\frac{M_E}{3 M_\odot} \right)^{1/3}. 
	\end{equation}
	Here, a = 1 A.U., and $M_\odot = 3 \times 10^{30} ~\text{kg}$ is the mass of Sun. Substituting the values, we obtain $ R_H = 0.01~\text{A.U.}$ The maximum final velocity for the DM particle can be written as,
	\begin{equation}\label{vf}
		v_f = \beta w 	
	\end{equation}
	where $\beta = 0.9957$, and  $w$ is the escape velocity as defined before. From Eq.\eqref{EF} and Eq. \eqref{vf} one obtains an upper limit on variable $x$ as,
	\begin{equation}\label{xmax}
	x_{max} = \frac{\beta^2}{f(\mu)^{N}}.
	\end{equation}
	It must be noted that in the original calculation \cite{Press}, the upper bound was motivated by considering an effective "scale height" beyond which incident flux was negligible. By considering Hill sphere of a body, we can have a physically inspired bound on the maximum incident energy. This will be significant change for planets like Jupiter and future calculation for other heavy bodies.\\
	
\begin{figure}[h]
			\centering
			\includegraphics[width = 8 cm, height = 7  cm]{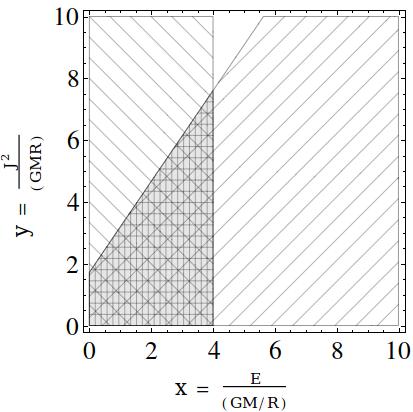}
			\caption{{\label{fig:capturable}The shaded region of the phase space is capturable for $x_{max} = 4$.}}
	\end{figure}

	The two constraints give the capturable phase-space as depicted in the Figure \ref{fig:capturable}. \\
	
	Using the recently obtained constraints as the limits of integration for \eqref{xyaccrate}, one obtains a total accretion rate of DM particles by Earth as, 
	\begin{equation}
	\resizebox{0.4\textwidth}{!}{$\mathcal{F} = \xi_\eta \sqrt{\pi}\left( \frac{3}{2}\right) ^{3/2} n_\chi \frac{w^4}{\bar{v}^3}R_E^2 g(\sigma_{\chi e}) \int_{0}^{x_{max}}dx \int_{0}^{1.42x+1.72} \exp\left(\frac{-3 w^2}{2\bar{v}^2} x \right)$}	
	\end{equation}
	
	Note that the mass of DM particle and the scattering cross section with electrons enter this equation through $x_{max}$ and $g(\sigma_{\chi e})$. 
	
\section{Heat Flux from annihilations/decays}

	The total number (N) of DM particles inside Earth follows the relation, 
	\begin{equation}\label{numberrate}
	 \dot{N} = \mathcal{F} - 2 \Gamma_A - \Gamma_D
	\end{equation}
	where $\mathcal{F}, \Gamma_A, \Gamma_D$ are the capture, annihilation, and decay rates respectively. The DM phase space distribution can be approximated by a thermal distribution depending on local gravitational potential \cite{Griest}.
	\begin{equation}
	 n(r) = n_0 e^{-m_\chi \phi(r)/k_b T_e}
	\end{equation}
	It is assumed that the DM will thermalize with electrons in Earth's core via scatterings. The temperature of DM can thus be assumed to be the that of electrons $T_e = 5000~ K$.
	We consider the cases of self annihilating DM and decaying DM separately. 
	\subsection{Decaying DM}
	For DM characterised by lifetime $\tau$, the decay rate is
	\begin{equation}
	\Gamma_D = \int d^3x~n(x)\left(\frac{1}{\tau}\right) = \frac{N}{\tau}
	\end{equation}
	We can solve the differential equation \eqref{numberrate} to obtain, 
	\begin{equation}
	N(t) = \mathcal{F}\tau (1 - \exp^{-t/\tau})
	\end{equation}
	Although we do not attain equilibrium, after a long time (few times the lifetime $\tau$), the decay rate of DM is given as, 
	\begin{equation}
	\Gamma_D = \frac{N(t\rightarrow \infty)}{\tau} \approx \mathcal{F}
	\end{equation}
	The heat produced due to decaying DM is thus, 
	\begin{equation}
	H_{dec} = \Gamma_D \times m_\chi = \mathcal{F}\times m_\chi
	\end{equation}
		
	\subsection{Self annihilating Dark Matter}
	For DM characterised by thermal-averaged annihilation cross section $\langle\sigma v\rangle$, the annihilaion rate is
	\begin{equation}
	\Gamma_A = \int d^3x~n^2(x) \langle\sigma v\rangle = \frac{\langle\sigma v\rangle V_2}{V_1^2} N^2 = \mathcal{A} N^2
	\end{equation}
	where 
	\begin{equation}
	V_j = 4\pi \int_{0}^{R_\odot}dr~r^2 \exp\left(-j \frac{m_\chi \phi(r)}{k_b T}\right)
	\end{equation}
	and
	\begin{equation}
	\mathcal{A} = \frac{\langle\sigma v\rangle V_2}{V_1^2}
	\end{equation}
	We can solve the differential equation \eqref{numberrate} to obtain, 
	\begin{equation}
	 N(t) = \sqrt{\frac{\mathcal{F}}{2\mathcal{A}}}\tanh\left(t\sqrt{2\mathcal{FA}}\right)
	\end{equation}
	
	Sufficiently long after the equilibrium time $\tau_{eq} = (2\mathcal{FA})^{-1/2}$, the annihilation rate is given as, 
	\begin{equation}
	\Gamma_A = \mathcal{A}N^2(t\rightarrow \infty) \approx \mathcal{F}/2
	\end{equation}
	The heat flux due to annihilating DM is thus, 
	\begin{equation}
	 H_A = \Gamma_a \times 2 m_\chi = \mathcal{F}\times m_\chi.
	\end{equation}
	
	In both the cases, one can see that the heat flux after equilibrium is given by $ \mathcal{F}\times m_\chi$. 
	
	\subsection*{Validity of Assumptions}
	
	One of the critical assumptions in the paper is that there is an equilibrium between capture and annihilaiton/decay of DM inside Earth. For annihilating DM with $\langle\sigma v\rangle_{\chi \chi} \geq 10^{-32} \text{cm}^3/s$ , it can be checked that this assumption is valid for majority of the parameter space in consideration. For decaying DM, it is required that the lifetime of DM be a few times less than the lifetime of Earth. \\ 
		
	It is assumed that the DM annihilates into SM particles which deposit their energy at core in form of heat. For majority of the parameter space i.e $m_{\chi} \leq 10$ MeV, the kinematically allowed annihilation/decay channels are electrons, photons, and neutrinos. Of these, a leptophilic DM would annihilate/decay into electrons at tree level and will be the most dominant channel. It annihilates/decays into photons and neutrinos at one-loop level, and amidst the two, neutrino channel is suppressed due to W-boson in the loop.\\
	
	For the rest of the parameter space, the possible tree level annihilation/decay channels also include muon and tau which eventually decay into electrons. Only the energy carried away by neutrinos is lost in such process. Assuming that on average the electrons and neutrinos carry equal energy, and taking branching fractions from \cite{pdg}, one can estimate that for $m_{\mu}<m_{\chi}<m_{tau}$ only 66\%, and for $m_{\tau}<m_{\chi}$ only 47\% energy is converted to heat. This modifies the curve only slightly in region already excluded by XENON10.\\
		
	As a model independent result, the parameter space in $\sigma_{\chi e} - m_\chi$ plane which produces more than 20 TW heat is shown in Fig.\ref{fig:mainres}. The exchange of energy between DM and electrons is most efficient when the DM mass is close to the mass of electrons. When this transfer is most efficient, even low scattering cross sections will result in capture of the incident particle. As the process becomes less efficient, we need larger cross sections for sufficient capture. This is the source of the V-shape of the curve. It can be interpreted in the following way: for a given mass of DM, if the scattering cross section is larger than a threshold value, too much DM will be captured by Earth which results in overheating. \\

\section{Results}
	We obtain constraint for low mass DM which has not been ruled out yet by direct detection experiments. Similar analysis has been performed for Moon's heat flux as measured by Apollo 15 and 17 \cite{Apollo}. Voyager-1 measured the internal heat flux of Jupiter and found anomalously large output \cite{Hanel}. As of now, there is no concrete explanation for this heat output. However if we assume that the future experiments and analysis can account for 80-90\% of this heat, we can put constraint using remainder of Jupiters heat. This is shown in the Fig. \ref{fig:earthmoonjup} below. \\
	\begin{figure}[h!]
				\centering
				\includegraphics[width = 8cm, height = 7  cm]{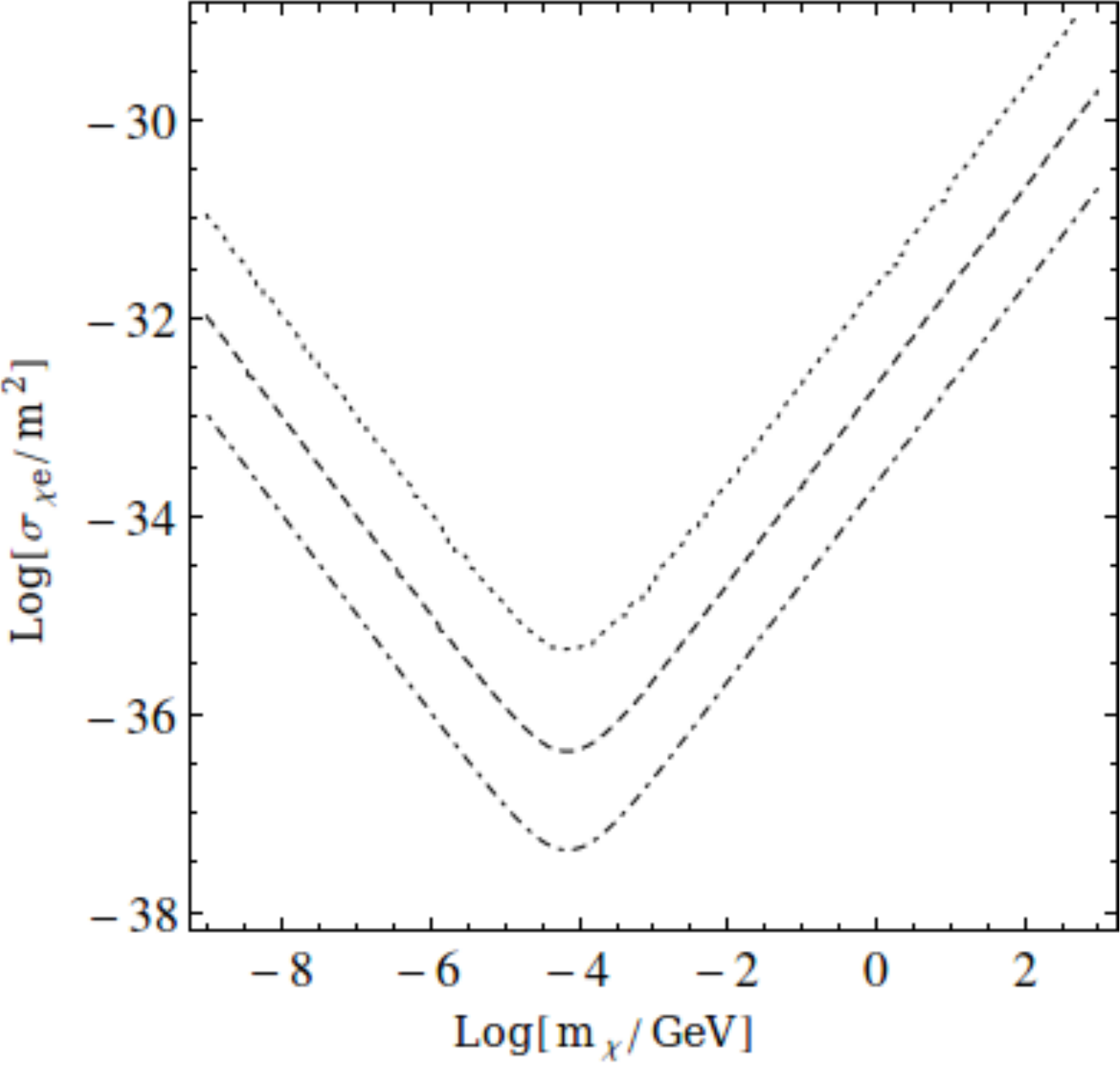}
				\caption{\label{fig:earthmoonjup} The constraints from Moon (dotted), Earth (dashed) and Jupiter (dot-dashed) are shown.}
	\end{figure}
	
	The bounds for Jupiter should not be taken too seriously for two reasons. First, the data is not reliable as the measurements were not \emph{in situ} as opposed to Earth and Moon. Secondly, the internal heat flux of Jupiter is so large, even if all the incident DM is captured, the heat produced is of the order of total internal heat. Hence a robust bound cannot be established. The most reliable and stringent bounds are established by Earth.\\

\begin{figure}[h!]
		\centering
		\includegraphics[width = 8cm, height = 7  cm]{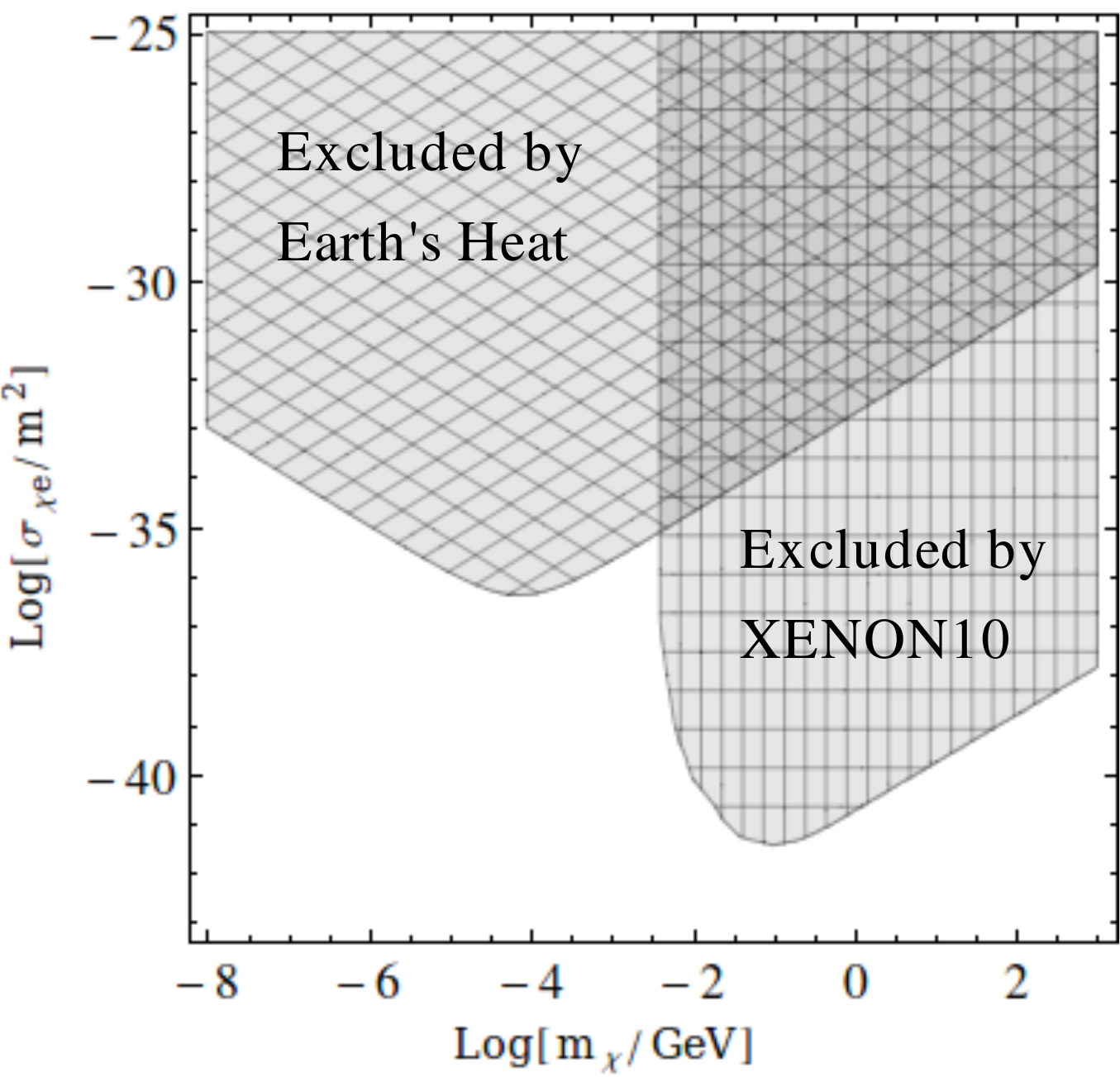}
		\caption{\label{fig:mainres}{The parameter space excluded from Xenon10 results and the result of this paper are shown together.}}
	\end{figure}

	In Fig.\ref{fig:mainres} we have shown the current bounds on sub-gev dark-matter as calculated in \cite{Essig2} and the new bounds from this paper. One can see that this parameter space for light leptophilic DM has not yet been ruled out by direct detection experiments.

\section{Conclusions}
	After analysing capture of local dark matter by Earth, we have found novel contraints on scattering cross section for leptopilic dark matter with electrons. We put new limit on the cross section of leptophilic DM in the mass range 1 eV - 3 MeV, which has not been probed by direct detection experiments yet. This result will be useful for testing already established and future models of leptophilic dark matter with mass in sub-MeV domain. \\
			
\bibliographystyle{apsrev4-1}
\bibliography{Resubmission.bib}

\end{document}